\documentclass[english,notitlepage,nofootinbib,superscriptaddress]{revtex4-1}
\pdfoutput=1
\usepackage[T1]{fontenc}
\usepackage[latin9]{inputenc}
\usepackage{color}
\usepackage{amsmath}
\usepackage{slashed}
\usepackage{graphicx}
\usepackage{esint}
\usepackage{url}
\usepackage{hyperref}

\makeatletter
\usepackage{afterpage}

\makeatother

\usepackage{babel}
\begin{document}

\title{Distinguishing between Dirac and Majorana neutrinos in the presence of general interactions}

\date{\today}

\author{Werner Rodejohann}
\affiliation{Max-Planck-Institut f\"ur Kernphysik, Postfach 103980, D-69029 Heidelberg,
Germany}
\author{Xun-Jie Xu}
\affiliation{Max-Planck-Institut f\"ur Kernphysik, Postfach 103980, D-69029 Heidelberg,
Germany}
\author{Carlos E. Yaguna }
\affiliation{Max-Planck-Institut f\"ur Kernphysik, Postfach 103980, D-69029 Heidelberg,
Germany}
\affiliation{Escuela de F\'{i}sica, Universidad Pedag\'ogica y Tecnol\'ogica de Colombia\\
Avenida Central del Norte, Tunja, Colombia}

\begin{abstract}
\noindent
We revisit the possibility of distinguishing between  Dirac and  Majorana neutrinos via neutrino-electron elastic scattering in the presence of all possible Lorentz-invariant interactions. Defining proper observables, certain regions of the parameter space can only be reached for Dirac neutrinos, but never for Majorana neutrinos, thus providing an alternative method to differentiate  these two possibilities. We first derive analytically and numerically the most general conditions that would allow to distinguish Dirac from Majorana neutrinos, both in the   relativistic  and non-relativistic cases.  Then, we apply these conditions to data on $\nu_\mu$-$e$ and $\bar{\nu}_e$-$e$ scatterings, from the CHARM-II and TEXONO  experiments, and find that they  are consistent with both types of neutrinos. Finally, we comment on  future prospects of this kind of tests.

\end{abstract}
\maketitle

\section{Introduction}

One of the most important questions in neutrino physics is whether neutrinos are Dirac or Majorana particles.
The most promising way to determine this may be through the observation of neutrinoless double beta decay
\cite{Rodejohann:2011mu}, though alternative possibilities are also often studied, see e.g.\   \cite{Cheng:1980tp,Kayser:1981nw,Shrock:1981cq,Shrock:1982sc,Rosen:1982pj,Garavaglia:1983wh,Barr:1987ht,Chhabra:1992be,Semikoz:1996up,Zralek:1997sa,Gaidaenko:1997cqa,Singh:2006ad,Menon:2008wa,Szafron:2009zz,Chen:2011hc,Szafron:2012mi,Dinh:2012qb,Alavi:2012np,Xing:2013ty,Barranco:2014cda,Frere:2015pma,Gluza:2016qqv,Dib:2016wge,Kim:2016bxw,Sobkow:2016zsg}.
The main difficulty in determining the Majorana character is that in theories with $V-A$ interactions such as the
Standard Model (SM), any observable difference between Dirac and Majorana neutrinos is always suppressed with
$(m_\nu/E)^2$, where $m_\nu$ is the neutrino mass and $E$ the energy scale of the process \cite{Kayser:1982br}. Examples are
here the suppression of the double beta  decay width with $m_\nu^2/(100 \, \rm MeV)^2$, the small relative
difference in the decay width of the $Z$ boson in Dirac or in Majorana neutrinos of order $(m_\nu/m_Z)^2$, or the suppression of neutrino-antineutrino oscillation probabilities with $(m_\nu/E)^2$.

However, if neutrinos have new interactions beyond the SM, the situation can be different --see  \cite{Sobkow:2016zsg} for a recent realization of this idea. Considering that massive neutrinos discovered via neutrino oscillations are regarded as an evidence of new physics beyond the SM, they may also be accompanied with new interactions. In an early work by Rosen \cite{Rosen:1982pj}, the most general Lorentz-invariant
form of neutrino-fermion interactions was assumed, including scalar, pseudo-scalar,
vector, axial-vector and tensor couplings. It was pointed out that
in elastic neutrino-electron scattering, due to the absence of vector and
tensor interactions in the Majorana case, the ratio of forward to
backward scattering cross sections, defined as $R_{\rho}$, could
be used to distinguish between Dirac and Majorana neutrinos. For Majorana
neutrinos, $R_{\rho}$ should be less than or equal to $2$, while
for Dirac neutrinos $R_{\rho}$ can be as large as 4. It is interesting to
compare this with $0\nu\beta\beta$. An observation of the process implies that neutrinos are Majorana particles.
Non-observation of the process implies either Dirac or Majorana particles, only additional input from other neutrino mass
approaches can possibly settle the situation \cite{Rodejohann:2011mu}.
In case of neutrino-electron scattering a measurement of $R_{\rho}>2$ would imply Dirac neutrinos, while $R_{\rho}\leq 2$
would imply either Dirac or Majorana particles.

In this paper, we will revisit Rosen's approach to distinguish Dirac from Majorana neutrinos.
We will in particular show that a detailed analysis of the problem reveals that even though
$R_{\rho}$ does have different allowed values for Dirac and Majorana neutrinos as described above, the situation is a bit more complicated. For example,
we will show that even if $R_{\rho}\leq2$, there is still a possibility
that Dirac and Majorana neutrinos can be distinguished in neutrino
scattering. Furthermore, the original work in Ref.\ \cite{Rosen:1982pj}
only applies to relativistic scattering with neutrino energies much
higher than the target particle mass. We will generalize the study
to the non-relativistic case where the target particle mass is not
negligible, which is important for instance for reactor
neutrino experiments.
We will also set limits on the strengths of the new interactions by using data from the
CHARM-II ($\nu_\mu$-$e$ scattering) and TEXONO ($\bar{\nu}_e$-$e$) experiments.

The paper is organized as follows: In the next section
we compute the neutrino-electron elastic scattering cross section in the presence of general new interactions. In
Sec.\ \ref{sec:new-bounds}, we first review Rosen's proposal to distinguish between Dirac and Majorana neutrinos in neutrino scattering and then generalize that analysis. In particular, two new ratios that can fully describe the differences between Dirac and Majorana neutrinos in relativistic scattering are introduced.  Next we further extend our analysis to the non-relativistic case, which is presented in Sec.\ \ref{sec:Non-relativ}. In Sec.\ \ref{sec:Experiment} we confront our criteria to experimental data, focusing on two experiments, CHARM-II and TEXONO, and we comment on future prospects. Finally, we summarize our work  in Sec.\ \ref{sec:Conclusion}.

\section{\label{sec:scat-in-SM} Neutrino-electron scattering with general interactions}
The most general Lorentz-invariant interaction of neutrinos with
charged leptons can be written as
\begin{equation}
{\cal L}\supset\frac{G_{F}}{\sqrt{2}}\sum_{a=S,P,V,A,T}\overline{\nu}\Gamma^{a}\nu\left[\overline{\ell}\Gamma^{a}(C_{a}+\overline{D}_{a}i\gamma^{5})\ell\right],\label{eq:dm}
\end{equation}
where $\Gamma^{a}$'s are the five possible independent combinations
of Dirac matrices defined as
\begin{equation}
\Gamma^{a}=\left\{I,i\gamma^{5},\gamma^{\mu},\gamma^{\mu}\gamma^{5},\sigma^{\mu\nu}\equiv\frac{i}{2}[\gamma^{\mu},\gamma^{\nu}]\right\}.\label{eq:dm-1}
\end{equation}
Here and henceforth we use the index $a=(S,\thinspace P,\thinspace V,\thinspace A,\thinspace T)$
to denote the above five cases which are usually  referred to as
scalar, pseudo-scalar, vector, axial-vector and tensor interactions,
respectively. The Lorentz indices in the first and second $\Gamma^{a}$
in Eq.\ (\ref{eq:dm}) should be properly contracted with each other (e.g.\
$\overline{\nu}\gamma^{\mu}\nu\left[\overline{\ell}\gamma_{\mu}(C_{V}+\overline{D}_{V}i\gamma^{5})\ell\right]$
for $a=V$). However, there could be two different ways to contract
in the tensor case, $g_{\mu\mu'}g_{\nu\nu'}\sigma^{\mu\nu}\sigma^{\mu'\nu'}$
and $\varepsilon_{\mu\nu\mu'\nu'}\sigma^{\mu\nu}\sigma^{\mu'\nu'}$.
Here we only take the former, since the latter
can be transformed to the former up to a redefinition of $C_{T}$
and $D_{T}$ (for more details, see Ref.\ \cite{Lindner:2016wff}).

Our convention is a little different from Rosen's original reference
\cite{Rosen:1982pj} in what regards the presence of $i$. We take the above convention
so that
\begin{equation}
\overline{\Gamma}^{a}\equiv\gamma^{0}(\Gamma^{a})^{\dagger}\gamma^{0}=\Gamma^{a}\,.\label{eq:dm-2}
\end{equation}
We also define
\begin{equation}
D_{a}\equiv\begin{cases}
\overline{D}_{a} & (a=S,\thinspace P,\thinspace T)\\
i\overline{D}_{a} & (a=V,\thinspace A)
\end{cases},\label{eq:dm-3}
\end{equation}
so that all the coupling constants $C_{a}$ and $D_{a}$ are real
numbers. In Ref.\ \cite{Dass:1984qc} the coupling constants of tensor
and scalar interactions are taken as complex numbers. Actually if
they are complex, then the interaction terms are not self-conjugate
so they should be added by their complex conjugates, which will make
$C_{a}$ and $D_{a}$ real. Here we do not assume CP conservation, while in Ref.\ \cite{Rosen:1982pj} for simplicity CP conservation is assumed so that $D_S=D_P=D_T=0$. Actually we find that the conclusions in Ref.\ \cite{Rosen:1982pj} still hold without CP conservation. 

It is well-known \cite{Rosen:1982pj,Dass:1984qc,Bergmann:1999rz}
that for Majorana neutrinos some coefficients should vanish:
\[
C_{V}=D_{V}=C_{T}=D_{T}=0\thinspace\thinspace{\rm (Majorana).}
\]
In the SM, the neutral current (NC) interaction is
\begin{equation}
{\cal L}_{{\rm NC}}=\frac{G_{F}}{\sqrt{2}}2\left[\overline{\nu}\gamma^{\mu}(g_{V}^{\nu}-g_{A}^{\nu}\gamma^{5})\nu\right]\left[\overline{\ell}\gamma_{\mu}(g_{V}^{\ell}-g_{A}^{\ell}\gamma^{5})\ell\right],\label{eq:dm-9}
\end{equation}
where
\begin{equation}
g_{V}^{\nu}=g_{A}^{\nu}=\frac{1}{2},\thinspace\thinspace g_{V}^{\ell}=-\frac{1}{2}+2s_{W}^{2},\thinspace g_{A}^{\ell}=-\frac{1}{2}.\label{eq:dm-10}
\end{equation}
The charged current (CC) interaction may also contribute to Eq.\ (\ref{eq:dm})
if the neutrino and the charged lepton
have the same flavor. To include the charged current contribution
one simply replaces $g_{V}^{\ell}\rightarrow g_{V}^{\ell}+1$ and $g_{A}^{\ell}\rightarrow g_{A}^{\ell}+1$ after a Fierz transformation.
From Eq.\ (\ref{eq:dm-9}) we can obtain the SM values of the couplings
in Eq.\ (\ref{eq:dm}) assuming Dirac neutrinos
\begin{equation}
C_{V}^{{\rm SM}}=2g_{V}^{\nu}g_{V}^{\ell},\thinspace D_{V}^{{\rm SM}}=-2g_{V}^{\nu}g_{A}^{\ell},\thinspace C_{A}^{{\rm SM}}=2g_{A}^{\nu}g_{A}^{\ell},\thinspace D_{A}^{{\rm SM}}=-2g_{A}^{\nu}g_{V}^{\ell}\thinspace\thinspace{\rm (Dirac),}\label{eq:dm-11}
\end{equation}
while the other couplings for $a=S,\thinspace P,\thinspace T$ should
be zero. If neutrinos are Majorana, i.e.\ $\nu$ in Eq.\ (\ref{eq:dm})
is a Majorana spinor, then according to Ref.\ \cite{Rosen:1982pj}
one should set $C_{V}$ and $D_{V}$ to zero and double $C_{A}$ and
$D_{A}$,
\begin{equation}
C_{V}^{{\rm SM}}=0,\thinspace D_{V}^{{\rm SM}}=0,\thinspace C_{A}^{{\rm SM}}=4g_{A}^{\nu}g_{A}^{\ell},\thinspace D_{A}^{{\rm SM}}=-4g_{A}^{\nu}g_{V}^{\ell}\thinspace\thinspace{\rm (Majorana).}\label{eq:dm-11-1}
\end{equation}
The cross section in both cases is the same in the SM (note that we neglect neutrino masses), as we will see next.

Now we can compute the cross section of elastic scattering of neutrinos
(antineutrinos) on charged leptons. Assuming the neutrino energy
is $E_{\nu}$ and the mass of charged leptons $M\ll E_{\nu}$, the
cross section in the laboratory frame is \textcolor{black}{\cite{Kayser:1979mj,Kingsley:1974kq}}\footnote{\textcolor{black}{In Refs.\ \cite{Kayser:1979mj,Kingsley:1974kq}
some couplings such as $D_{S}$, $D_{P}$, $D_{T}$ are set to zero.
}To get a more general cross section with non-zero \textcolor{black}{$D_{S}$,
$D_{P}$ and $D_{T}$, we use} FeynCalc \cite{Shtabovenko:2016sxi,Mertig:1990an}
and Package-X \cite{Patel:2015tea} to compute the cross section again,
which is consistent with the result in \textcolor{black}{\cite{Kayser:1979mj,Kingsley:1974kq}
in the zero limit of $D_{S}$, $D_{P}$ and $D_{T}$.}}
\begin{eqnarray}
\frac{d\sigma}{dT}(\nu+\ell) & = & \frac{G_{F}^{2}M}{2\pi}\left[A+2B\left(1-\frac{T}{E_{\nu}}\right)+C\left(1-\frac{T}{E_{\nu}}\right)^{2}\right],\label{eq:dm-4}
\end{eqnarray}
\begin{eqnarray}
\frac{d\sigma}{dT}(\overline{\nu}+\ell) & = & \frac{G_{F}^{2}M}{2\pi}\left[C+2B\left(1-\frac{T}{E_{\nu}}\right)+A\left(1-\frac{T}{E_{\nu}}\right)^{2}\right],\label{eq:dm-5}
\end{eqnarray}
where we have defined
\begin{equation}
A\equiv\frac{1}{4}\left(C_{A}-D_{A}+C_{V}-D_{V}\right){}^{2}+\frac{1}{2}C_{P}C_{T}+\frac{1}{8}(C_{P}^{2}+C_{S}^{2}+D_{P}^{2}+D_{S}^{2})-\frac{1}{2}C_{S}C_{T}+C_{T}^{2}+\frac{1}{2}D_{P}D_{T}-\frac{1}{2}D_{S}D_{T}+D_{T}^{2}\,,\label{eq:dm-6}
\end{equation}
\begin{equation}
B\equiv-\frac{1}{8}\left(C_{P}^{2}+C_{S}^{2}+D_{P}^{2}+D_{S}^{2}\right)+C_{T}^{2}+D_{T}^{2}\,,\label{eq:dm-7}
\end{equation}
\begin{equation}
C\equiv\frac{1}{4}\left(C_{A}+D_{A}-C_{V}-D_{V}\right){}^{2}-\frac{1}{2}C_{P}C_{T}+\frac{1}{8}(C_{P}^{2}+C_{S}^{2}+D_{P}^{2}+D_{S}^{2})+\frac{1}{2}C_{T}C_{S}+C_{T}^{2}-\frac{1}{2}D_{P}D_{T}+\frac{1}{2}D_{S}D_{T}+D_{T}^{2}\,,\label{eq:dm-8}
\end{equation}
and $T$ is the recoil energy of the charged lepton.
Note that Eqs.\ (\ref{eq:dm-4}) and
(\ref{eq:dm-5}) are derived under the assumption that 
 the incoming neutrinos or anti-neutrinos are left-handed or right-handed, respectively. However this assumption requires that the neutrinos are produced via the CC interaction in the SM, such as in beta decay, pion decay etc. In principle, one could also consider new neutrino interactions in the production of neutrinos, but for this one needs to introduce some other new interactions with new independent parameters.  Our calculations assume incoming 
left-handed neutrinos or right-handed anti-neutrinos.

It is interesting to note that the vector and axial-vector interactions do not interfere with the other interactions (scalar, pseudo-scalar, tensor) in the cross sections (\ref{eq:dm-4}) and
(\ref{eq:dm-5}). This is due to the property that $\Gamma^V=\gamma^{\mu}$ and $\Gamma^A=\gamma^{\mu}\gamma^{5}$ have odd numbers of gamma matrices ($\gamma^5$ is a combination of four gamma matrices) while in $\Gamma^S=I$, $\Gamma^P=i\gamma^{5}$, $\Gamma^T=\sigma^{\mu\nu}$ the number is even. When computing the cross section 
involving the interference of $\Gamma^a $ and $\Gamma^b $, the relevant part of the squared amplitude is proportional to $\slashed{p} P_R \Gamma^a \slashed{k} \Gamma^b P_L$, where $p$/$k$ is the incoming/outgoing momentum of the neutrino and $P_{L/R}=(1\mp\gamma^5)/2$. If $\Gamma^a$ and $\Gamma^b$ have even and odd 
numbers, respectively,  in the number of gamma matrices, then $P_R$ can be put in front of $P_L$ according to the commutation relations of gamma matrices, resulting to zero. 
If both are even or odd, then the relevant part of the amplitude does not vanish. 
Therefore, interactions containing odd gamma matrices ($V$ and $A$) do not interfere with interactions containing an even number of gamma matrices ($S$, $P$ and $T$) but 
interference terms do appear among ($V$, $A$) as well as among ($S$, $P$, $T$).


Since the $A$, $B$, $C$ terms in Eqs.\ (\ref{eq:dm-4}) and
(\ref{eq:dm-5}) have different energy dependencies, they can be extracted from  scattering data by fitting the  event distributions with respect to the recoil energy. Besides, one could also combine the neutrino and antineutrino channels to obtain a better determination of $A$, $B$, $C$, according to Eqs.\ (\ref{eq:dm-4}) and (\ref{eq:dm-5}).

If the neutrino interactions are described by the  SM, the cross section has the same value for  Dirac and Majorana neutrinos,  as one can check by using Eqs.\ (\ref{eq:dm-11}) and (\ref{eq:dm-11-1}) to compute $A$, $B$, $C$ defined in Eqs.\,(\ref{eq:dm-6})-(\ref{eq:dm-8}). The SM values\footnote{Note that due to the RG running we take different values of $s_{W}^{2}$ at different scattering energies.
} of $A$, $B$ and $C$ are
\begin{equation}
(A,\thinspace B,\thinspace C)^{{\rm SM}}=\left((1-2s_{W}^{2})^{2},\thinspace0,\thinspace4s_{W}^{4}\right)\thinspace\thinspace{\rm (NC\thinspace only),}\label{eq:dm-12}
\end{equation}
if only the NC interaction is present in the scattering (e.g.\ for $\nu_{\mu}+e^{-}\rightarrow\nu_{\mu}+e^{-}$).
If the neutrino has the same flavor as the charged lepton (e.g.\ for  $\nu_{e}+e^{-}\rightarrow\nu_{e}+e^{-}$),
then the CC interaction makes an additional contribution so that
\begin{equation}
(A,\thinspace B,\thinspace C)^{{\rm SM}}=\left((1+2s_{W}^{2})^{2},\thinspace0,\thinspace4s_{W}^{4}\right)\thinspace\thinspace{\rm (NC+CC).}\label{eq:dm-12-1}
\end{equation}
However, when the other interactions  in Eq.\ (\ref{eq:dm}) exist, the cross sections are expected to be different
for the Dirac and Majorana cases.

\section{\label{sec:new-bounds} Distinguishing Dirac from Majorana neutrinos}

In Ref.\ \cite{Rosen:1982pj}
Rosen proposed a measurable ratio
\begin{equation}
R_{\rho}\equiv\frac{2(A+2B+C)}{A+C}\label{eq:dm-13}
\end{equation}
to distinguish between Dirac and Majorana neutrinos. One could determine this ratio experimentally by either binning the neutrino or antineutrino cross section in at least three bins, or by evaluating the ratio of forward $(T=0)$ and backward $(T=E_\nu)$  scattering of the sum of neutrino and antineutrino scattering.
The ratio $R_\rho$ has different bounds for the two cases:
\begin{eqnarray}
0 & \leq & R_{\rho}\leq4 \thinspace({\rm Dirac}),\label{eq:dm-14}\\
0 & \leq & R_{\rho}\leq2 \thinspace({\rm Majorana}).\label{eq:dm-15}
\end{eqnarray}
If $R_{\rho}$ is found to be larger than 2 and less than or equal
to 4 then it implies neutrinos are Dirac particles. If it would
lie in the range of $[0,2]$ then one can not distinguish between
Dirac and Majorana neutrinos.

Note that the SM value of $R_{\rho}$ is 2.
Majorana neutrinos only allow a
downward deviation while Dirac neutrinos allow both downward and upward
deviations.

The bounds (\ref{eq:dm-14}) and (\ref{eq:dm-15}) proposed by Rosen
imply that the parameter space of $(A,\thinspace B,\thinspace C)$
in the Dirac case is larger than that in the Majorana case. Some values
of $(A,\thinspace B,\thinspace C)$ that can be reached by Dirac neutrinos
are not allowed in the Majorana case, which is used as a criterion
to determine the nature of neutrinos. However, Eqs.\ (\ref{eq:dm-14})
and (\ref{eq:dm-15}) do not fully cover the difference. There
are some other parts of the parameter space that also enable us to exclude Majorana neutrinos.  Next,
we will examine in full generality the parameter space of $(A,\thinspace B,\thinspace C)$ to find the most general conditions that allows to discriminate Dirac from Majorana neutrinos.

\begin{figure}
\centering

\includegraphics[width=8cm]{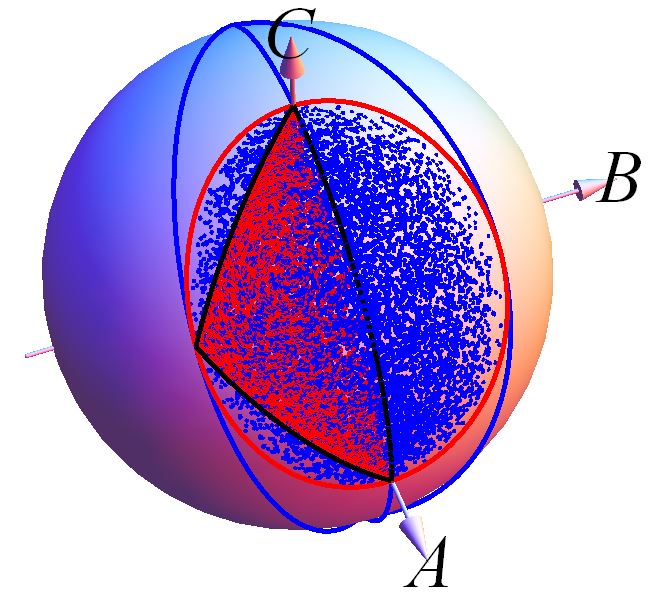}

\caption{\label{fig:ball} Allowed values of normalized $(A,\thinspace B,\thinspace C)$
assuming Dirac (blue points) or Majorana (red points) neutrinos. The
blue curves represents Rosen's original results ($R_{\rho}=0$, $2$ and $4$ from left to right, respectively) while the actually
allowed ranges are smaller, surrounded by the red curves (for Dirac
neutrinos) or black curves (for Majorana neutrinos).}
\end{figure}

Since the dependence of $(A,\thinspace B,\thinspace C)$ on the couplings $(C_{a},\thinspace D_{a})$ is very complicated, we first try a random
scan to numerically find out the boundary of the space.
That is, we randomly generate arbitrary values of $(C_{a},\thinspace D_{a})$
with $a=S,\thinspace P,\thinspace V,\thinspace A$ and $T$, except
that for Majorana neutrinos we set $C_{V}=D_{V}=C_{T}=D_{T}=0$.
Then we compute the corresponding values of $(A,\thinspace B,\thinspace C)$
according to Eqs.\ (\ref{eq:dm-6})-(\ref{eq:dm-8}).
Note that for any allowed value of $(A,\thinspace B,\thinspace C)$,
$(rA,\thinspace rB,\thinspace rC)$ is also allowed for any positive
$r$ since it just corresponds to a rescaling of $(C_{a},\thinspace D_{a})$
by a factor of $\sqrt{r}$. Therefore for simplicity, we can normalize
$(A,\thinspace B,\thinspace C)$ to show the allowed region. In Fig.\ \ref{fig:ball}
we show the normalized values of $10^{4}$ samples generated by randomly
choosing values of $(C_{a},\thinspace D_{a})$ in $[-1,1]$. The red
points are for the Majorana case and blue points for Dirac, which
are confined by the black and red curves respectively. The bounds
(\ref{eq:dm-14}) and (\ref{eq:dm-15}) are marked by the blue curves,
with $R_{\rho}=0$, $2$ and $4$ from left to right, respectively.
The original criterion proposed by Rosen is that if $(A,\thinspace B,\thinspace C)$
is measured between the central and the right blue curves then neutrinos
should be Dirac particles. If $(A,\thinspace B,\thinspace C)$
is measured such that one ends up between the central and
the left blue curves, both Dirac and Majorana are possible.

However, the allowed regions are smaller, as we can see from the blue
and red points. Furthermore, the difference between the Dirac and
Majorana cases is not simply that the allowed region of the latter
is half of the former. The Majorana region does not fully cover the
left half of the Dirac region, thus revealing  new criteria to distinguish
between Dirac and Majorana neutrinos. For example, if $B<0$ then
from the parameter $R_{\rho}$ one can not determine the nature of
neutrinos since $R_{\rho}$ is always less than 2. But if $(A,\thinspace B,\thinspace C)$
falls into the small region between the left boundary of Majorana
and the left boundary of Dirac, one can still draw the conclusion
that neutrinos are Dirac particles.

\begin{figure}
\centering

\includegraphics[width=6cm]{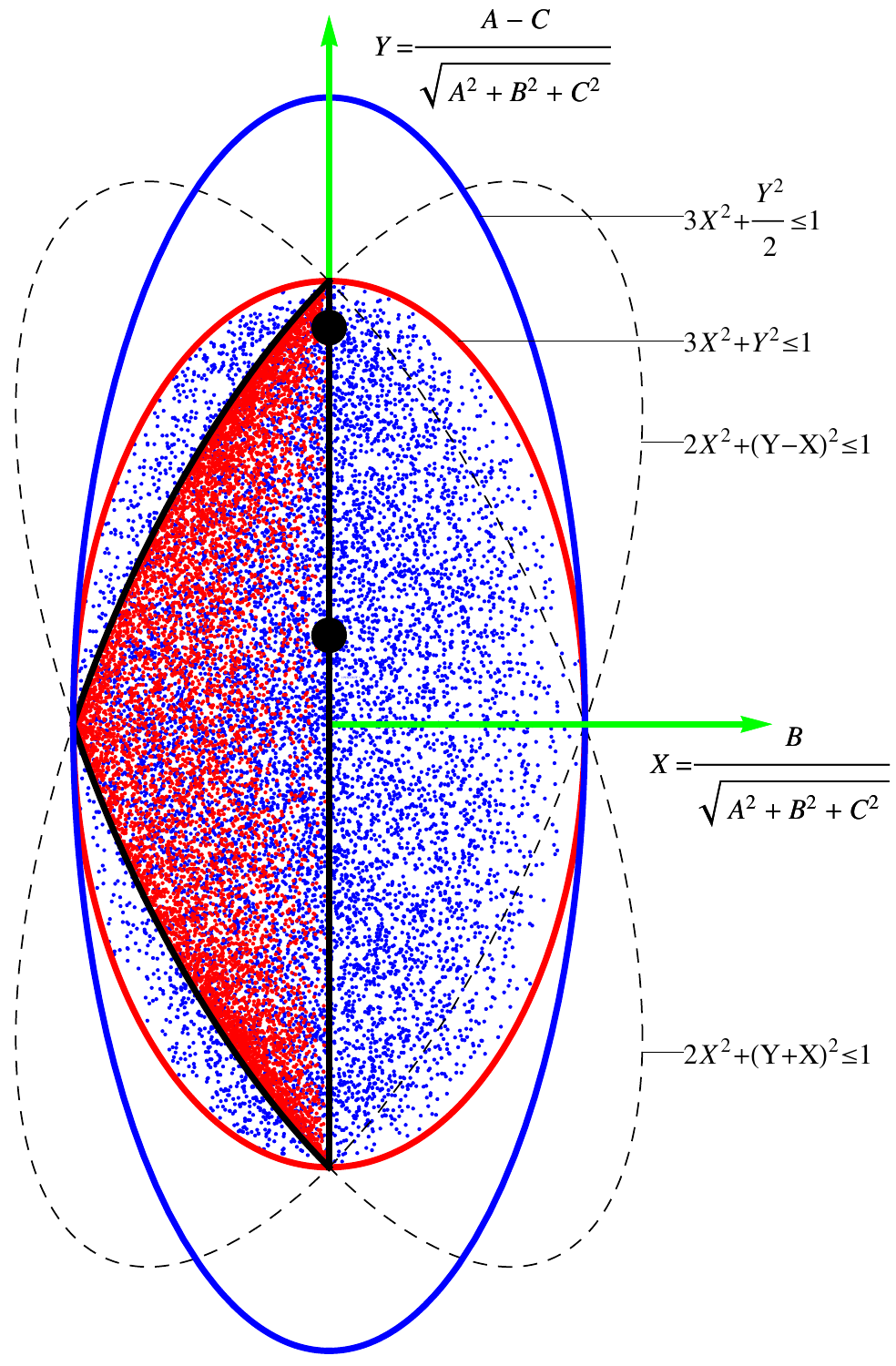}

\caption{\label{fig:eggs} Various bounds on $X$ and $Y$ defined in Eq.\ (\ref{eq:dm-16}),
converted from Fig.\ \ref{fig:ball} to a two-dimensional form. The two black dots in the
middle indicate the SM values, the lower and upper one corresponding
to NC  (e.g.\ $\nu_{\mu}+e^{-}$ scattering) and NC+CC (e.g.\ $\nu_{e}+e^{-}$),
respectively. }
\end{figure}

Since the scale factor of $(A,\thinspace B,\thinspace C)$ is not
important here, we define two ratios $X$ and $Y$:
\begin{equation}
X\equiv\frac{B}{R}\,,\thinspace Y\equiv\frac{A-C}{R}\,,\thinspace\label{eq:dm-16}
\end{equation}
where $R$ is the normalization factor,
\begin{equation}
R\equiv\sqrt{A^{2}+B^{2}+C^{2}}\,.\label{eq:dm-43}
\end{equation}
The two ratios $X$ and $Y$ suffice to describe the difference
between the Dirac and Majorana parameter spaces. Using $X$ and $Y$,
we can present the parameter spaces in a two-dimensional form, as shown in Fig.\ \ref{fig:eggs}.
The Dirac bound is marked by the red ellipse; the analytic
expression (derived in the appendix) to describe it is
\begin{equation}
{\rm Dirac\thinspace bound:\quad}3X^{2}+Y^{2}\leq1\,.\label{eq:dm-17}
\end{equation}
The Majorana bound consists of parts of the two black dashed ellipses and
a central vertical line, described by the expressions (also derived in the appendix)
\begin{equation}
{\rm Majorana\thinspace bound:}\quad2X^{2}+(Y\pm X)^{2}\leq1\quad{\rm and}\quad X\leq0\,.\label{eq:dm-18}
\end{equation}
The region bounded by the black solid line fulfills those three criteria.
Besides, we also show Rosen's Dirac bound ($0\leq R_{\rho}\leq4$)
by the blue ellipse, which in this formulation is $3X^{2}+Y^{2}/2\leq1$.
For Rosen's Majorana bound, only the left half of the blue ellipse is allowed, i.e.\  $\sqrt{1/3-Y^{2}/6}\leq X\leq0$.
The two black dots on the central vertical line represent the SM values,
without or with the CC interaction included, according to Eqs.\ (\ref{eq:dm-12})
and (\ref{eq:dm-12-1}), respectively.

\section{\label{sec:Non-relativ}Non-relativistic scattering}

\afterpage{%
\begin{figure}[h]
\centering

\includegraphics[height=0.75\paperheight]{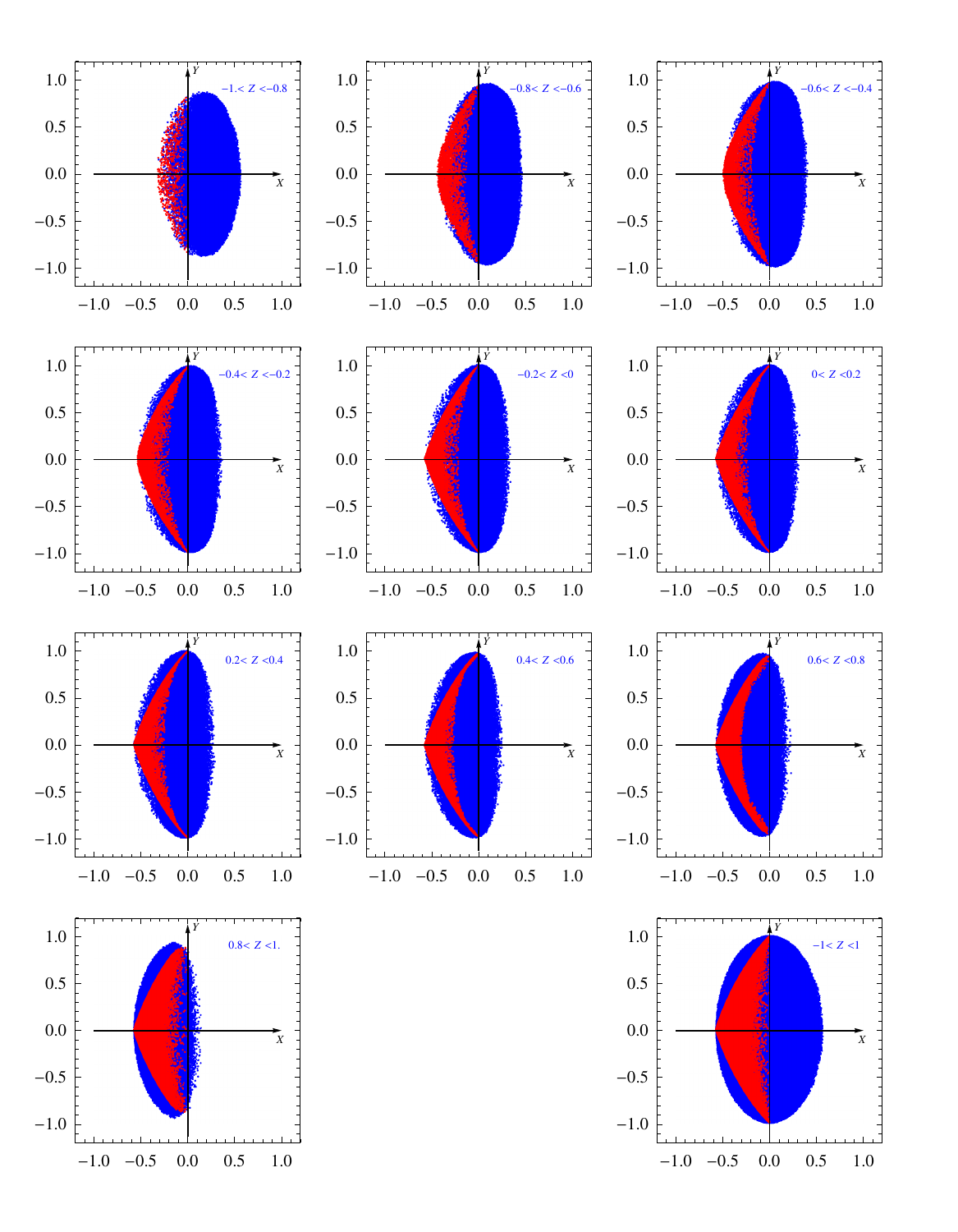}

\caption{\label{fig:XYZ} Dirac and Majorana bounds on $(X,\thinspace Y,\thinspace Z)$
in non-relativistic scattering. For different values of $Z$, allowed
regions are shown with blue (Dirac) and red (Majorana) points. The
last plot (right bottom) shows the combination of all possible values
of $Z$, which reproduces the result from Fig.\ \ref{fig:eggs}.}
\end{figure}

\clearpage
}

So far we have only considered the case with neutrino energies much
higher than charged lepton masses ($E_{\nu}\gg M$) so that all particles
are assumed approximately massless. For low energy neutrino scattering,
this approximation may not hold so we need to take the mass into account.
For example, in reactor neutrino experiments, neutrino energies are
typically at the MeV scale\footnote{A typical reactor neutrino flux peaks roughly near 1 MeV and then
decreases quickly above 2 MeV. However, the uncertainty of the flux
below 2 MeV is quite large \cite{Kopeikin:2012zz,Schreckenbach:1985ep}; often
the events are selected from 3 MeV to 8 MeV.} which are not much higher than the electron mass, 0.511 MeV.  In this
section, we consider the more general case with the charged lepton
masses taken into account.

The result for the cross section of elastic scattering of neutrinos with massive
charged leptons is
\begin{eqnarray}
\frac{d\sigma}{dT}(\nu+\ell) & = & \frac{G_{F}{}^{2}M}{2\pi}\left[A+2B\left(1-\frac{T}{E_{\nu}}\right)+C\left(1-\frac{T}{E_{\nu}}\right)^{2}+D\frac{MT}{4E_{\nu}^{2}}\right],\label{eq:dm-4-1}
\end{eqnarray}
\begin{eqnarray}
\frac{d\sigma}{dT}(\overline{\nu}+\ell) & = & \frac{G_{F}{}^{2}M}{2\pi}\left[C+2B\left(1-\frac{T}{E_{\nu}}\right)+A\left(1-\frac{T}{E_{\nu}}\right)^{2}+D\frac{MT}{4E_{\nu}^{2}}\right],\label{eq:dm-5-1}
\end{eqnarray}
where
\begin{equation}
D\equiv\left(C_{A}-D_{V}\right){}^{2}-\left(C_{V}-D_{A}\right){}^{2} +C_{S}^{2}-4C_{T}^{2}+D_{P}^{2}-4D_{T}^{2}.\label{eq:dm-38}
\end{equation}
Note that the $A$, $B$, $C$ terms remain the same as the relativistic
case {[}cf.\ Eqs.\ (\ref{eq:dm-4})-(\ref{eq:dm-8}){]} while the $D$
term is the only additional term that contains the contribution of
nonzero charged lepton mass. It is the same for both the neutrino
and antineutrino channels. In the SM, the value of $D$ is given
by\footnote{Note that in the relativistic case, $D$ is not measurable but it
does not mean $D$ is zero. So for all the equations or conclusions
in this section, to get their relativistic limit one needs to set $M\rightarrow0$, not $D\rightarrow0$.}
\begin{equation}
D^{{\rm SM}}=1-(1+4s_{W}^{2})^{2}\,.\label{eq:dm-40}
\end{equation}
From Eqs.\ (\ref{eq:dm-4-1}) and (\ref{eq:dm-5-1}) one can deduce that there
are four parameters $(A,\thinspace B,\thinspace C,\thinspace D)$
that could be measured in non-relativistic neutrino scattering. If
there were measurements of $(A,\thinspace B,\thinspace C,\thinspace D)$,
then again their values could be used to distinguish between Dirac
and Majorana neutrinos. The difference between the relativistic
and non-relativistic cases is that in the former we have only three
measurable quantities $(A,\thinspace B,\thinspace C)$ while in the
latter we have four, $(A,\thinspace B,\thinspace C,\thinspace D)$.
However, in both cases $(A,\thinspace B,\thinspace C)$ are subject
to the same constraint shown in Fig.\ \ref{fig:ball}, which implies
our previous bounds on $(A,\thinspace B,\thinspace C)$ in relativistic
scattering also apply here.
Actually if the scattering is quasi-relativistic, e.g.\ for
reactor neutrinos scattering on electrons,
$M/E_{\nu}$ is small\footnote{But not negligibly small so that one still has to use Eqs.\ (\ref{eq:dm-4-1})
or (\ref{eq:dm-5-1}) instead of Eqs.\ (\ref{eq:dm-4}) or (\ref{eq:dm-5}).}, and the measurement of $D$ would be the most inaccurate among the four
quantities. So for such experiments, the most important quantities
would still be $(A,\thinspace B,\thinspace C)$. In case $D$ were known, a ratio such as
\begin{equation}
R_\sigma = \frac{A + 2 B + C - D}{A+C}
\end{equation}
could be studied. For Dirac neutrinos  $R_\sigma$ lies within $[-4,4]$, while for Majorana neutrinos between $[-4,3]$.

Next, we shall study the Dirac and Majorana bounds in the presence
of four measurable quantities $(A,\thinspace B,\thinspace C,\thinspace D)$.
Similar to the relativistic case, the normalization of $(A,\thinspace B,\thinspace C,\thinspace D)$
is not important  so we only care about their ratios. From the four
quantities, we can define at most three independent ratios, we choose here
\begin{equation}
(X,\thinspace Y,\thinspace Z)\equiv\left(\frac{B}{\sqrt{A^{2}+B^{2}+C^{2}}},\thinspace\frac{A-C}{\sqrt{A^{2}+B^{2}+C^{2}}},\thinspace\frac{D}{\sqrt{A^{2}+B^{2}+C^{2}+D^{2}}}\right).\label{eq:dm-39}
\end{equation}
Note that $D^{2}$ appears in the denominator of the definition of
$Z$, but not in $X$ and $Y$. In principle we could also put $D^{2}$
in the definitions of $X$ and $Y$, but we choose the current form
because we want to keep the same definitions of $X$ and $Y$ as the
relativistic case. As we have just mentioned, the previous bounds
in the relativistic case on $(A,\thinspace B,\thinspace C)$ can also
be used in the non-relativistic case and might probably be the most
important bounds if the scattering is quasi-relativistic, so we prefer
to define the first two ratios independent of $D$, which enable us
to apply the relativistic bounds directly to the non-relativistic
case.

With the three ratios, the problem essentially becomes finding the
allowed regions in the $X$-$Y$-$Z$ space for Dirac and Majorana
neutrinos. Finding the boundaries analytically as for the
relativistic case in the previous section turns out to be much more complicated and difficult than the 2-dimensional problem that lead to Fig.\ \ref{fig:eggs}. Though the
projections of these regions in some special directions can be analytically
computed, they can not provide a full description of the allowed regions.
So instead, we will use a numerical method to find the allowed
regions. We generate $10^{6}$ samples with random values of $(C_{a},\thinspace D_{a})$
in $[-1,1]$ and then compute the corresponding $(X,\thinspace Y,\thinspace Z)$
according to Eqs.\ (\ref{eq:dm-6})-(\ref{eq:dm-8}), (\ref{eq:dm-38})
and (\ref{eq:dm-39}).  Since the allowed regions are 3-dimensional
objects, we display their 2-dimensional projection with various $Z$ values.
 The results are presented in Fig.\ \ref{fig:XYZ},
where blue points are for Dirac neutrinos and red for Majorana. For each fixed value of $Z$, the section shows the allowed values
of $(X,\thinspace Y)$. As
one may expect, combining the results of all possible $Z$ values should
lead to a result analogous to that shown in Fig.\ \ref{fig:eggs}.

\section{\label{sec:Experiment}Experimental constraints}

As we have derived different bounds for Dirac and Majorana neutrinos
in the presence of general interactions, we would like to confront
these bounds with current experimental constraints. Possible experimental
constraints may directly come from elastic neutrino scattering experiments
such as  CHARM \cite{Dorenbosch:1988is}, CHARM-II \cite{Vilain:1993kd,Vilain:1994qy},
LAMPF \cite{Allen:1992qe}, MINER$\nu$A \cite{Park:2015eqa}, LSND \cite{Auerbach:2001wg}
and TEXONO \cite{Deniz:2009mu}, or  from other  experiments sensitive
to new neutrino interactions, such as LEP or observations of
atmospheric, reactor, solar, and accelerator neutrinos in neutrino
oscillation experiments.

A complete  analysis including all possible experimental constraints
is beyond the scope of this paper. Instead we will only focus on the
strongest constraints from elastic neutrino scattering experiments\footnote{Note that for the widely-studied Non-Standard Interaction (NSI) there
has been a comprehensive study on the various constraints from present
and future experiments \cite{Davidson:2003ha}. It turns out that
for $\varepsilon_{\mu\mu}^{e}$ and $\varepsilon_{ee}^{e}$, the current
strongest constraints are from elastic neutrino scattering experiments
(see table 2 in \cite{Davidson:2003ha}). Therefore we expect that
the strongest constraints on more general new interactions may also
come from such experiments. This is one of the reasons that we only
focus on elastic neutrino scattering experiments in this work.}, which provide direct measurements on $A$, $B$, $C$ and $D$ (if
the scattering is relativistic, then $D$ can not be measured) according to Eqs.\,(\ref{eq:dm-4})
and (\ref{eq:dm-5}) or \,(\ref{eq:dm-4-1}) and (\ref{eq:dm-5-1}).

Since our discussion in this section may refer to different channels
such as $\nu_{\mu}+e$ or $\bar\nu_{e}+e$, we would like to distinguish
them by adding subscripts of the corresponding lepton flavors. If
the flavors of the neutrino and charged lepton are $\alpha$ and $\beta$
respectively $(\alpha,\thinspace\beta=e,\thinspace\mu,\thinspace\tau)$
then the corresponding $(A,\thinspace B,\thinspace C,\thinspace D)$
is denoted by $(A_{\alpha\beta},\thinspace B_{\alpha\beta},\thinspace C_{\alpha\beta},\thinspace D_{\alpha\beta})$,
and $(X,\thinspace Y)$ by $(X_{\alpha\beta},\thinspace Y_{\alpha\beta})$.

The most interesting and promising channel to achieve a high precision
of measurement are $(\alpha,\thinspace\beta)=(\mu,\thinspace e)$
or $(e,\thinspace e)$.  In this paper, we will not
include all the elastic neutrino scattering experiments but only
two  representative experiments, CHARM-II (for the $\mu e$ channel)
and TEXONO (for the $ee$ channel) as they may provide the strongest
constraints\footnote{Actually their measurements of $s_{W}^{2}$ may be a good indicator
of the precision, $s_{W}^{2}=0.2324\pm0.0083$ in CHARM-II \cite{Vilain:1994qy},
$0.249\pm0.063$ in LAMPF \cite{Allen:1992qe}, $0.248\pm0.051$ in
LSND \cite{Auerbach:2001wg} and $0.251\pm0.031\pm0.024$ in TEXONO
\cite{Deniz:2009mu}. From $s_{W}^{2}$ we can expect that CHARM-II
and TEXONO provide the strongest constraints on new interactions in
the corresponding channels.}.

\begin{figure}
\centering

\includegraphics[width=8cm]{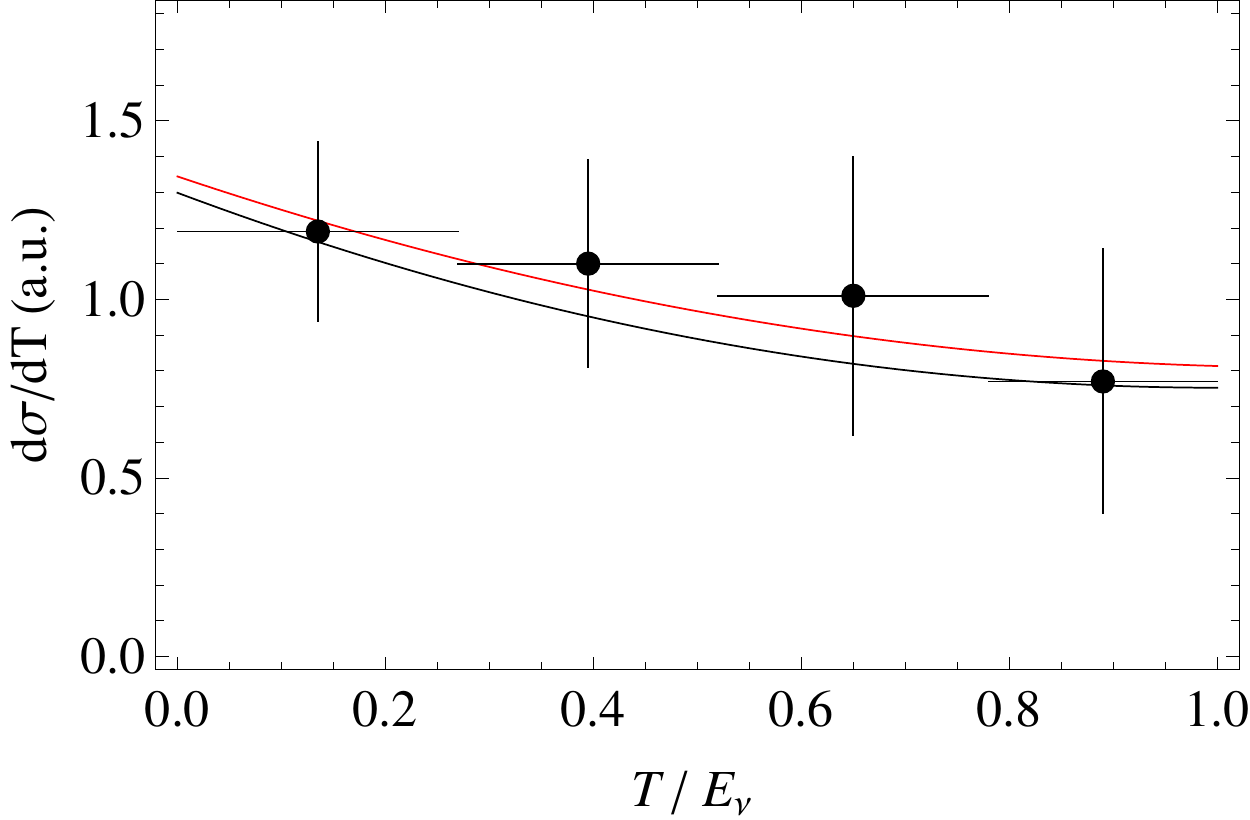}\,\includegraphics[width=8cm]{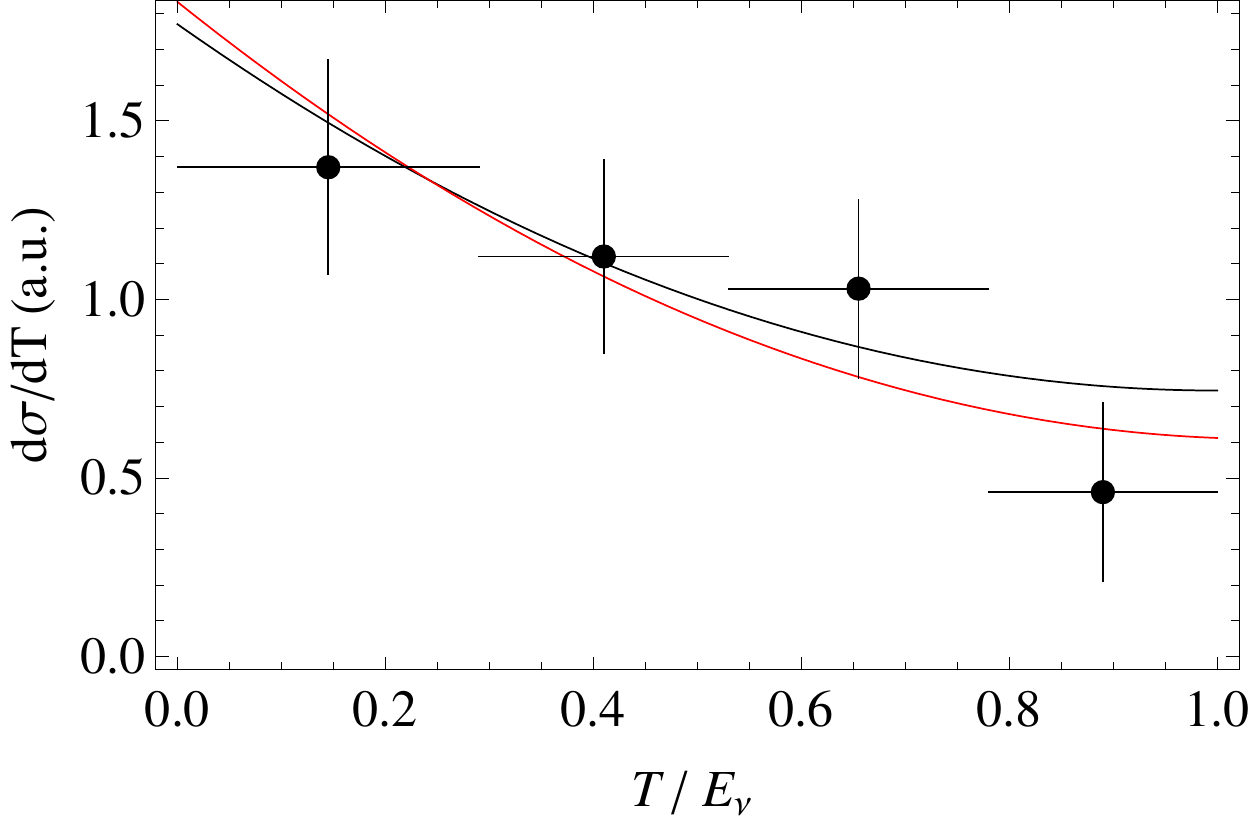}

\caption{\label{fig:fit-curve} Differential cross sections of $\nu_{\mu}+e^{-}$
(left) and $\overline{\nu}_{\mu}+e^{-}$ (right) measured in CHARM-II. The black points  represent the data of CHARM-II measurement,
taken from \cite{Vilain:1993kd}. The black and red  curves are the
SM prediction and the best fit, respectively.}
\end{figure}

The CHARM-II collaboration has measured both $\nu_{\mu}$ and $\overline{\nu}_{\mu}$
neutrinos scattering on electrons with a mean neutrino energy at 20
GeV (i.e.\ the scattering is highly relativistic). The unfolded differential
cross sections from this measurement have been published in \cite{Vilain:1993kd},
from which it is quite straightforward to perform a $\chi^{2}$-fit
to obtain the constraint on $(A_{\mu e},\thinspace B_{\mu e},\thinspace C_{\mu e})$,
including both $\nu_{\mu}$ and $\overline{\nu}_{\mu}$ data. We consider the following $\chi^2$-function:
\begin{equation}
\chi^{2}(A_{\mu e},\thinspace B_{\mu e},\thinspace C_{\mu e})=\sum_{i=T\thinspace{\rm bins}}\frac{\left[(\frac{d\sigma}{dT})_{i}-s_{i}\right]^{2}}{\sigma_{s,i}^{2}}+(\nu_{\mu}\rightarrow\overline{\nu}_{\mu})\,,\label{eq:dm-37}
\end{equation}
where $s_{i}$ and $\sigma_{s,i}$ represent the measured differential
cross section of $\nu_{\mu}+e^{-}$ and its uncertainty respectively.
 The antineutrino part is also included in the $\chi^{2}$-fit. The data of $s_{i}$
and $\sigma_{s,i}$, taken from \cite{Vilain:1993kd}, is shown in
Fig.\ \ref{fig:fit-curve}.

\begin{figure}
\centering

\includegraphics[height=9cm]{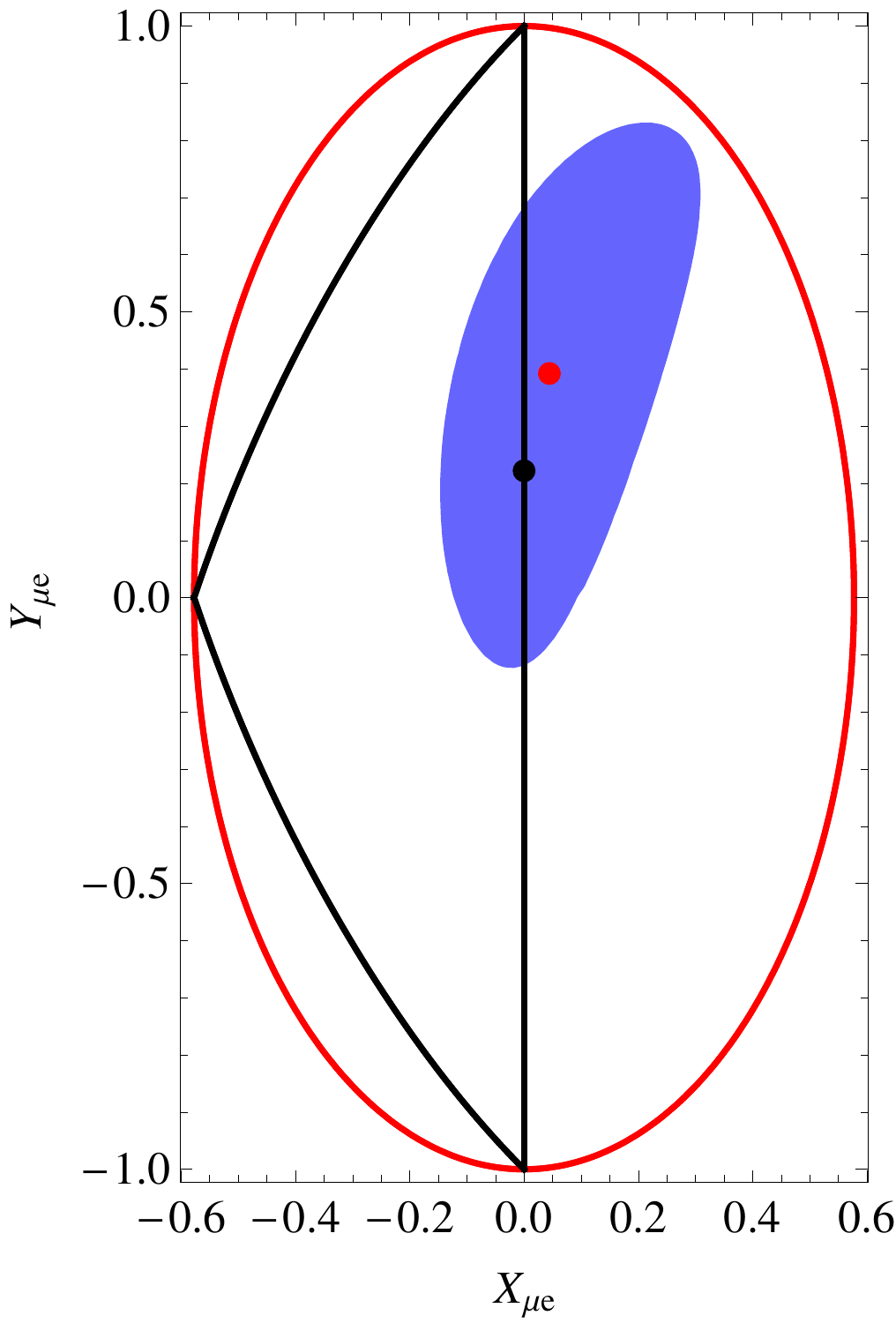}$\quad\quad$\includegraphics[height=9cm]{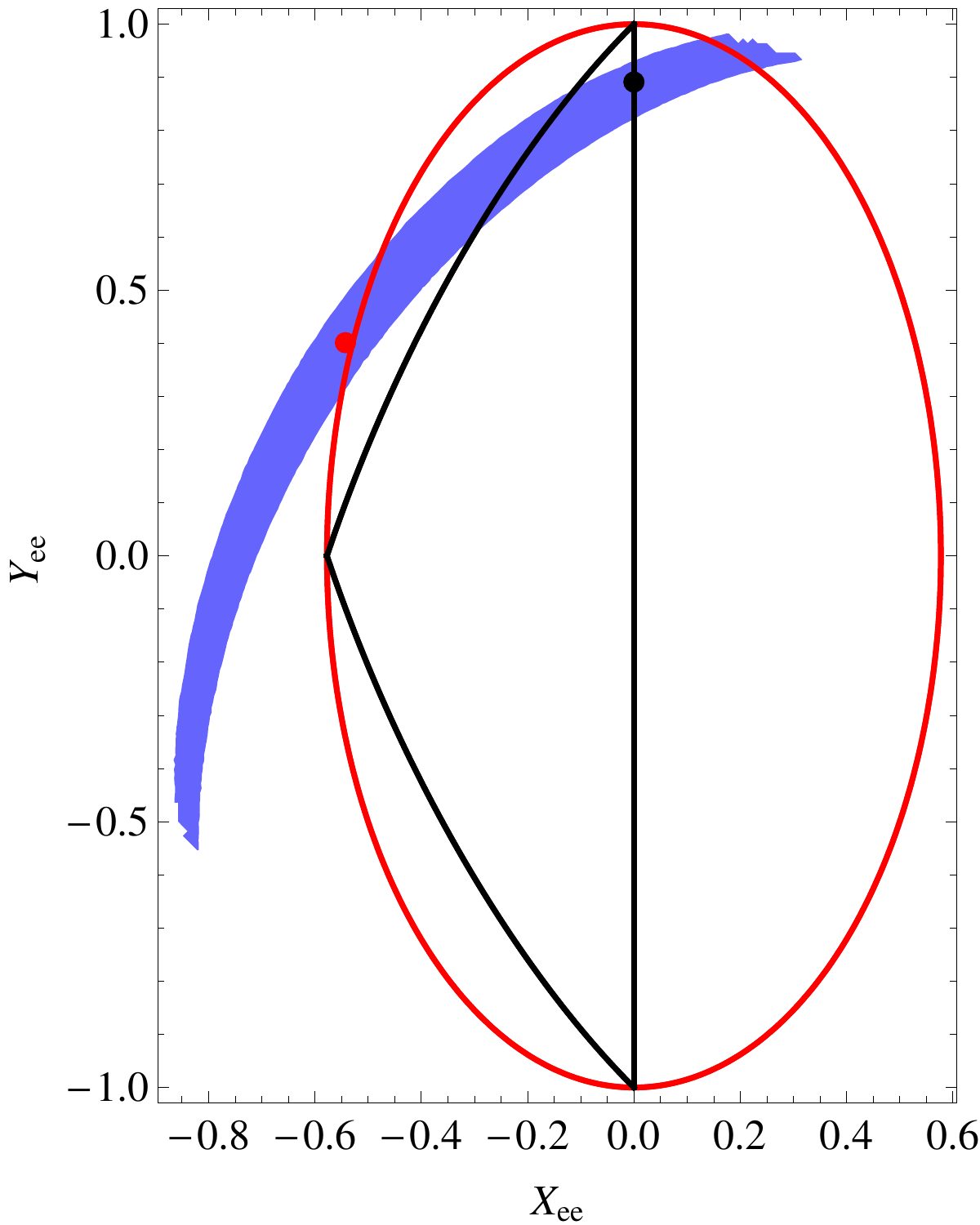}

\caption{\label{fig:eggs-1} Constraints from elastic neutrino scattering experiments
in the $\mu e$ channel (left) and the $ee$ channel (right). The
blue regions correspond to 90\% C.L.\ constraints from CHARM-II \cite{Vilain:1993kd}
($\mu e$ channel) and TEXONO ($ee$ channel).  The black and red
points represent the SM values and the best fit. respectively.
The black and red curves are the Majorana and Dirac bounds in Eqs.\ (\ref{eq:dm-17})
and (\ref{eq:dm-18}).}
\end{figure}

For any given $(A_{\mu e},\thinspace B_{\mu e},\thinspace C_{\mu e})$
we can compute the corresponding $\chi^{2}$-value according to Eq.\ (\ref{eq:dm-37}).
However, to distinguish between Dirac and Majorana neutrinos, what
actually concerns us is $(X_{\mu e},\thinspace Y_{\mu e})$. One can
convert the $\chi^{2}$-fit on $(A_{\mu e},\thinspace B_{\mu e},\thinspace C_{\mu e})$
to the corresponding fit on $(X_{\mu e},\thinspace Y_{\mu e})$ according
to Eq.\ (\ref{eq:dm-16}) if the normalization factor $R_{\mu e}$
is known. In this work, for simplicity, we fix $R_{\mu e}$ at the
SM value.  The result is presented in the left panel of Fig.\ \ref{fig:eggs-1}
where the 90\% C.L.\ constraint is shown by the blue region. The black
and red points represent  the SM value and the best fit, respectively.

From the left panel of Fig.\ \ref{fig:eggs-1} we can see that large
deviations of $(X_{\mu e},\thinspace Y_{\mu e})$ from the the SM
value (the black point) are still allowed by the CHARM-II constraint
(the blue region), which covers both the Dirac and Majorana regions.
Both the best fit (the red point) and the SM value (black point) are well compatible with data, which can be directly understood from the red and
black curves in Fig.\ \ref{fig:fit-curve}.

Currently a large part of the blue region (including the best fit)
locates in the $X_{\mu e}>0$ region, which can only be reached if
neutrinos are Dirac. Though the constraint is too weak\footnote{Here we have only used the data of unfolded differential cross sections
of CHARM-II \cite{Vilain:1993kd}. The constraint obtained in this
way is weaker than the actual constraint from the original event spectrum,
because the unfolding process would inevitably lead to a loss of information
of the original spectrum.  Since the original event spectrum is
not accessible to us, we have to use the unfolded data instead.} to show any preference for Dirac or Majorana neutrinos,  it
illustrates how  future measurements might  enable us to determine the nature of neutrinos: such data may prefer a region inaccessible to Majorana neutrinos. \\

The TEXONO experiment \cite{Deniz:2009mu} has measured elastic $\overline{\nu}_{e}+e^{-}$ reactor neutrino scattering with a CsI(Tl) crystal detector. The
scattering events are selected from 3 MeV to 8 MeV. In this experiment,
the electron mass is not negligible and should be taken into account.
So we shall use the non-relativistic formula (\ref{eq:dm-5-1}). The
event rates in each recoil energy bin can be estimated by
\begin{equation}
N_{i}=\int_{T_{i}}^{T_{i}+\Delta T}dT\int_{0}^{8\thinspace{\rm MeV}}dE_{\nu}\Phi(E_{\nu})\frac{d\sigma}{dT}(T,E_{\nu})\,,\label{eq:dm-41}
\end{equation}
where $\Delta T$ is the bin width and $\Phi(E_{\nu})$ is the reactor
neutrino flux. We use the following $\chi^{2}$-function to fit the
parameters in the cross section,
\begin{equation}
\chi^{2}(A_{ee},\thinspace B_{ee},\thinspace C_{ee},\thinspace D_{ee})=\sum_{i=T\thinspace{\rm bins}}\frac{\left[N_{i}-N_{i}^{0}\right]^{2}}{\sigma_{N,i}^{2}}\,,\label{eq:dm-42}
\end{equation}
where $N_{i}^{0}$ and $\sigma_{N,i}$ are the observed event rates
and the corresponding uncertainties, respectively. We take $\Phi(E_{\nu})$,
$N_{i}^{0}$ and $\sigma_{N,i}$ all from \cite{Deniz:2009mu}. The result is presented in the right panel
of Fig.\ \ref{fig:eggs-1} where to convert the $\chi^{2}$-fit on
$(A_{ee},\thinspace B_{ee},\thinspace C_{ee},\thinspace D_{ee})$
to $(X_{ee},\thinspace Y_{ee})$ we have fixed the normalization factors
$R_{ee}$ and $D_{ee}$ to their SM values.

\begin{figure}
\centering

\includegraphics[width=10cm]{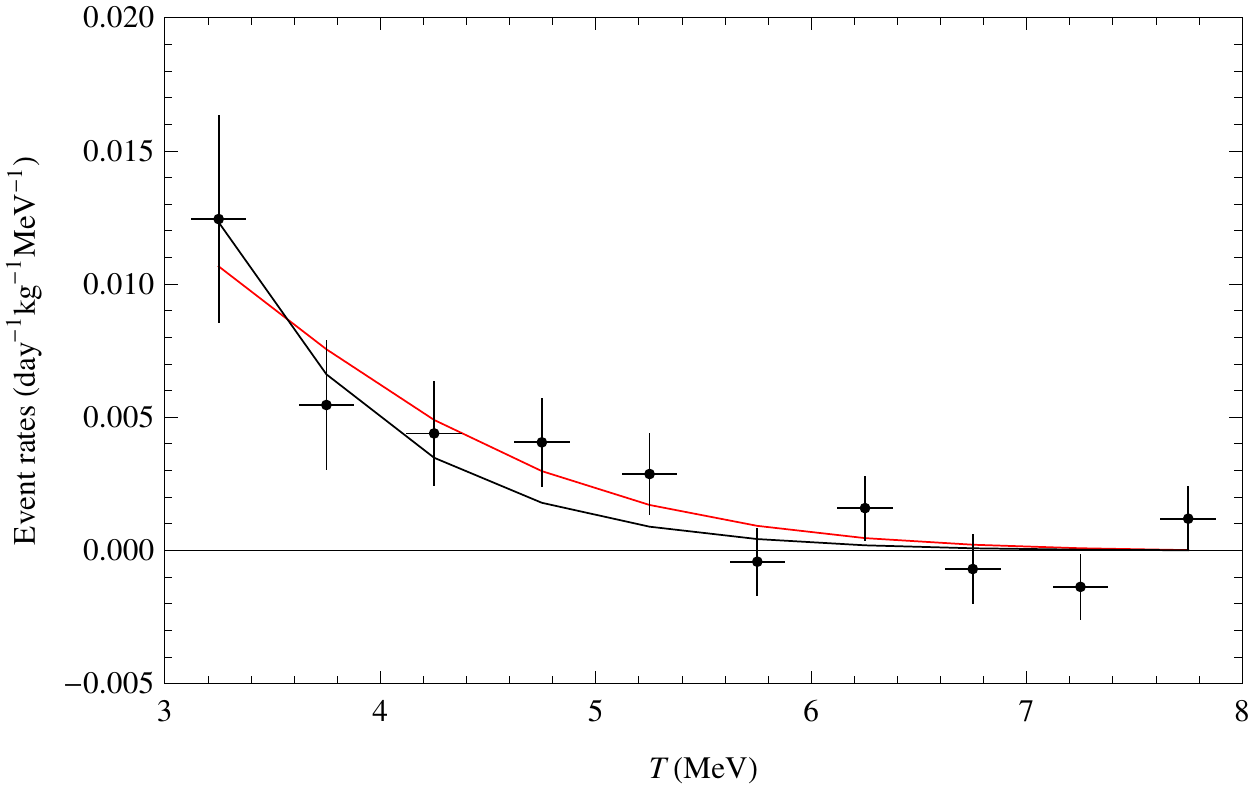}\,

\caption{\label{fig:fit-curve-1} Event rates of $\overline{\nu}_{e}+e^{-}$
elastic scattering in the TEXONO experiment. The black points  represent
the measured data, taken from \cite{Deniz:2009mu}. The black and
red  curves are the SM prediction and the best fit, respectively.}
\end{figure}

As is shown in Fig.\ \ref{fig:eggs-1}, the current constraint from
TEXONO not only allows both the Dirac and Majorana cases, but also
allows the parameters to be outside the Dirac bound. Actually the current
best fit (the red point) is just on the borderline of this case. Indeed, the red curve in Fig.\ \ref{fig:fit-curve-1}
which is generated with the parameters at this point, shows directly
that it can fit the TEXONO data very well. If the scattering parameters turn out to
lie outside  the Dirac bound, then this would  imply
exotic new physics that can not be described by the most general Lorentz
invariant interactions in Eq.\ (\ref{eq:dm}).

\begin{figure}
\centering

\includegraphics[width=8cm]{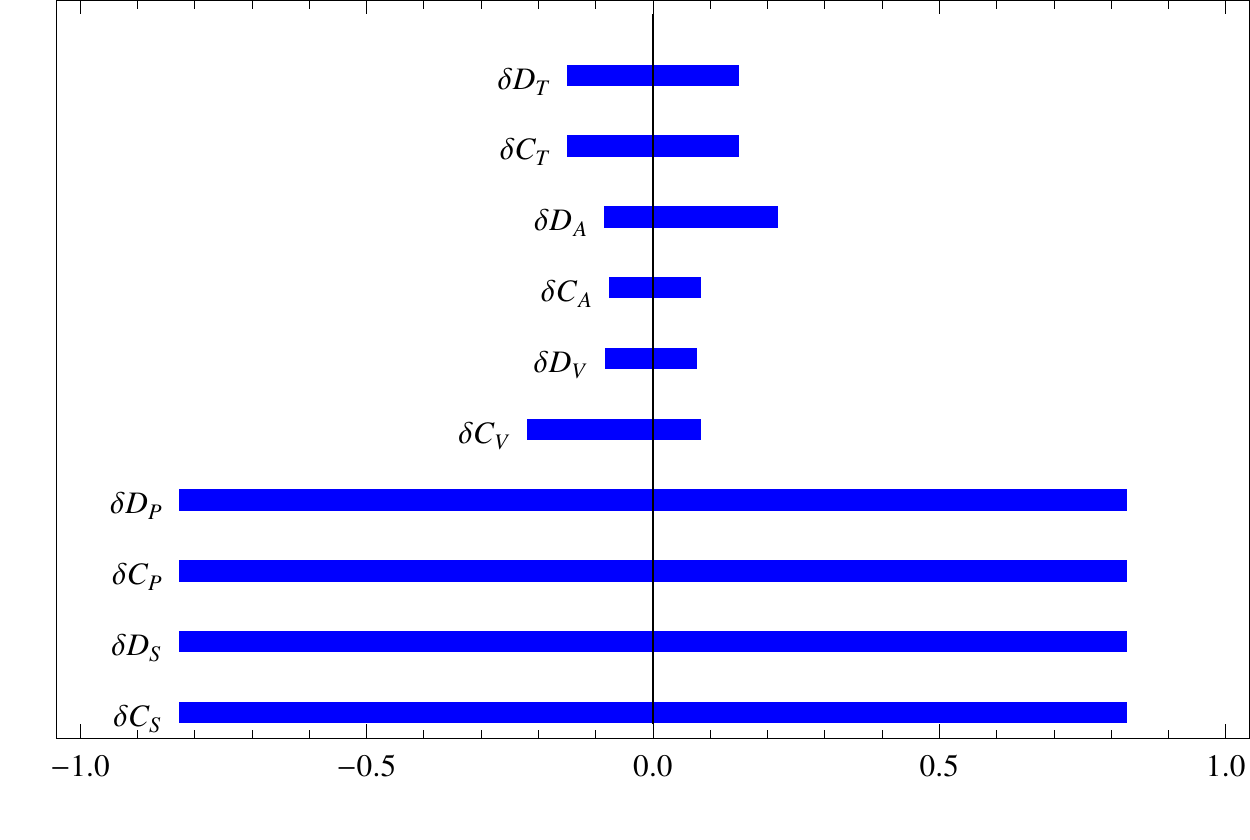}\
\includegraphics[width=8cm]{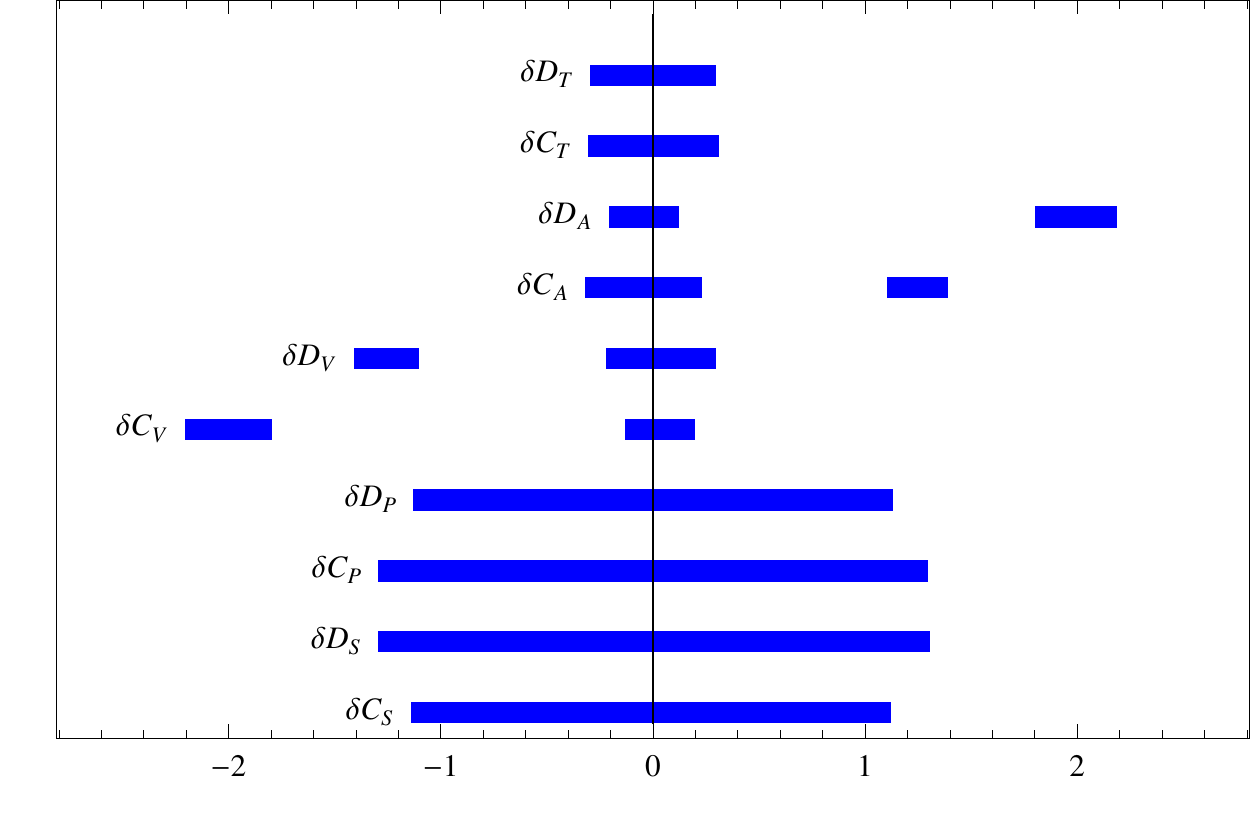}

\caption{\label{fig:deltaCD}
Constraints on $\delta C_a\equiv C_a-C^{\rm{SM}}_a$ and $\delta D_a\equiv D_a-D^{\rm{SM}}_a$ from one-parameter fitting of CHARM-II (left panel)
and TEXONO (right panel). Blue bars represent $90\%$ C.L.\ allowed values for $\delta C_a$ and $\delta D_a$. There are two local minima for
$C_{A,V}$ and $D_{A,V}$ in the fit of TEXONO. The slightly more minimal
global minimum is around $-2$ for $C_V$, $0$ for
$D_V$, $0$ for $C_A$ and $+2$ for $D_A$.
In the SM, $C_{V,A}$ and $D_{V,A}$ are non-zero for Dirac neutrinos, while only $C_{A}$ and $D_{A}$ are non-zero for Majorana neutrinos. The plot assumes Dirac neutrinos. }
\end{figure}

From the fits it is straightforward to obtain the current limits on the fundamental parameters $C_a$ and $D_a$ in Eqs.\ \eqref{eq:dm} and \eqref{eq:dm-3}. The result is shown in Fig.\ \ref{fig:deltaCD}. The fit assumes Dirac neutrinos, for Majorana neutrinos the results look very similar. As one can see, small values of tensor interactions and  small  departures from the SM values are allowed.  Scalar and pseudo-scalar interactions are weakly constrained.
Recently Ref.\ \cite{Sevda:2016otj}
published constraints on tensor and scalar interactions from
TEXONO data, which as we have checked is consistent with our result.

\begin{figure}
\centering

\includegraphics[height=9cm]{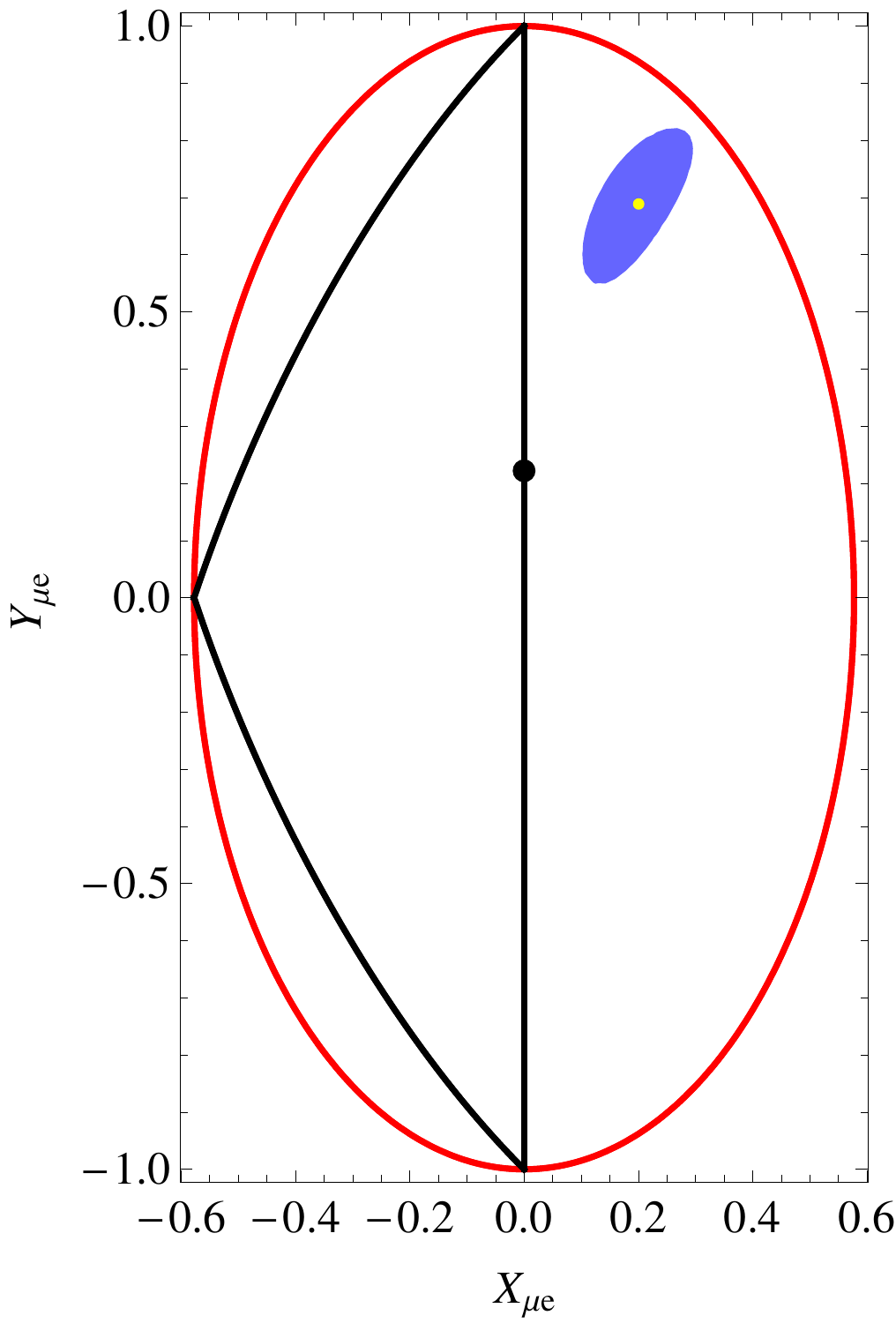}$\quad\quad$\includegraphics[height=9cm]{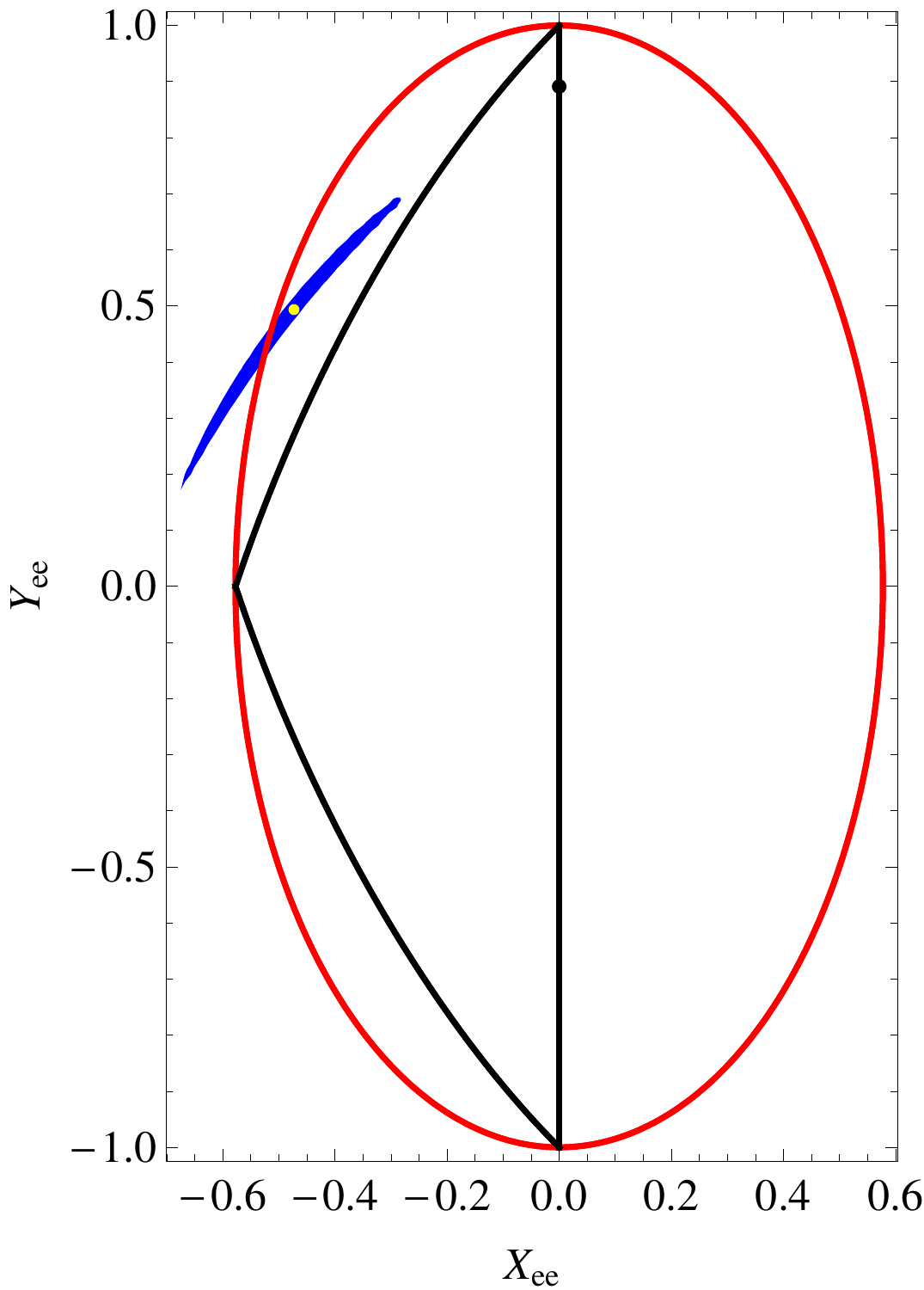}

\caption{\label{fig:eggs-future} Future constraints from elastic neutrino
scattering experiments in the $\mu e$ channel (left) and the $ee$
channel (right). We assume the actual values of $(X,\thinspace Y)$
are located at the yellow points, and the uncertainties of the future
experiments are reduced by a factor of $3$ or $4$ with respect to
CHARM-II or TEXONO, respectively. Other details are the same as Fig.\ \ref{fig:eggs-1}.}
\end{figure}

Since the current constraints from CHARM-II and TEXONO allow both
the Dirac and Majorana cases, we would like to investigate whether future
experiments with  improved sensitivities could distinguish
between them. This possibility will crucially depend on how far away the
actual values of $(X,\thinspace Y)$ are from the SM values. For
illustration, we choose two points which are still in the $90\%$
C.L.\ bounds of CHARM-II and TEXONO, but significantly deviate from the
SM values. Then we assume that future experiments
have  improved sensitivities so that compared to CHARM-II
or TEXONO, the uncertainties of measurement would be reduced by a
factor of 3 or 4, respectively. In Fig.\ \ref{fig:eggs-future}, we
show the $90\%$ C.L.\ bounds of such hypothetical experiments (the blue regions).
We can see that if the actual
values (the yellow points) deviate from the SM values (the black points)
significantly, then such experiments could exclude Majorana neutrinos
at more than $90\%$ C.L.

\section{\label{sec:Conclusion}Conclusion}

If neutrinos have new interactions beyond the SM, then their Dirac/Majorana
nature could have observable differences in neutrino scattering, which are  not suppressed by neutrino masses. We have performed an exhaustive study
on all possible criteria that could be used to distinguish between
Dirac and Majorana neutrinos in this context.

We have computed the cross sections of elastic neutrino-fermion scattering in the presence of the  most
general interactions including scalar, pseudo-scalar, vector, axial-vector
and tensor interactions. The result is given by Eqs.\ (\ref{eq:dm-4}),
(\ref{eq:dm-5}) for the relativistic case and it implies that there
are at most three independent scattering parameters $(A,\thinspace B,\thinspace C)$
that could be measured in this case. These parameters are subject
to certain bounds which depend on the Dirac/Majorana nature of neutrinos,
as shown in Fig.\ \ref{fig:ball} or Fig.\ \ref{fig:eggs} in terms
of two ratios $(X,\thinspace Y)$ defined from $(A,\thinspace B,\thinspace C)$.
Therefore the Dirac and Majorana bounds in Fig.\ \ref{fig:ball}
or Fig.\ \ref{fig:eggs} {[}for analytic expressions, see Eqs.\ (\ref{eq:dm-17})
and (\ref{eq:dm-18}){]} could be used to determine the nature of
neutrinos. If the parameters measured in neutrino scattering experiments
are out of the Majorana bound but inside the Dirac bound, then
neutrinos are Dirac particles.

As for the non-relativistic case, we find there is only one additional
term in the cross sections {[}cf.\ Eqs.\ (\ref{eq:dm-4-1}), (\ref{eq:dm-5-1}){]}.
This extends the three scattering parameters $(A,\thinspace B,\thinspace C)$
in the relativistic case to four parameters $(A,\thinspace B,\thinspace C,\thinspace D)$
and the two ratios $(X,\thinspace Y)$ to three, $(X,\thinspace Y,\thinspace Z)$,
with their explicit expressions given by Eq.\ (\ref{eq:dm-38}) and
Eq.\ (\ref{eq:dm-39}). The corresponding Dirac and Majorana bounds
are numerically found and presented in Fig.\ \ref{fig:XYZ}. Note
that the bounds found for the relativistic case, i.e.\ without $D$,
also apply to the non-relativistic case.

Currently the constraints from two neutrino scattering experiments, CHARM-II and TEXONO, are well consistent with both Dirac and Majorana neutrinos
as shown in Fig.\ \ref{fig:eggs-1}. Nevertheless they are able to
effectively constrain the ratios $(X,\thinspace Y)$. If in the future
neutrino scattering experiments would achieve much higher precision,
deviations from the SM might be observed, from which one might be able
to determine the nature of neutrinos. So far we have not included
measurements from other experiments, which if combined may give
stronger constraints.

Our analysis shows that there is still room for exciting new physics in the neutrino sector, and searches for new neutrino interactions are surely of large interest, as they can contribute to the exciting question of whether neutrinos are Dirac or Majorana particles.

\acknowledgments

This work is supported by the DFG with grant RO 2516/6-1 in
the Heisenberg program (WR) and with grant RO 2516/5-1 (CY).

\appendix*
\section{Proof}
Here we shall analytically derive
the Dirac bound (\ref{eq:dm-17}) and the Majorana bound (\ref{eq:dm-18}).
In the Dirac case, we write $(A,\thinspace B,\thinspace C)$ in the
following form
\begin{eqnarray}
A & = & E^{2}+G^{2}+H^{2}-K\,,\label{eq:dm-19}\\
B & = & -G^{2}+H^{2}\,,\label{eq:dm-20}\\
C & = & F^{2}+G^{2}+H^{2}+K\,,\label{eq:dm-21}
\end{eqnarray}
where
\begin{eqnarray}
 &  & E^{2}\equiv\frac{1}{4}\left(C_{A}-D_{A}+C_{V}-D_{V}\right){}^{2}\,,\quad F^{2}\equiv\frac{1}{4}\left(C_{A}+D_{A}-C_{V}-D_{V}\right){}^{2}\,,\label{eq:dm-22}\\
 &  & G^{2}\equiv\frac{1}{8}(C_{P}^{2}+C_{S}^{2}+D_{P}^{2}+D_{S}^{2})\,,\quad H^{2}\equiv C_{T}^{2}+D_{T}^{2}\,,\label{eq:dm-23}\\
 &  & K\equiv\frac{1}{2}C_{T}(C_{S}-C_{P})+\frac{1}{2}D_{T}(D_{S}-D_{P})\,.\label{eq:dm-24}
\end{eqnarray}
Note that here $(E,\thinspace F,\thinspace G,\thinspace H)$ can be
treated as free parameters without any constraints among them, i.e.\
for any values of $(E,\thinspace F,\thinspace G,\thinspace H)$ one
can always find the corresponding $(C_{a},\thinspace D_{a})$ to generate
them. After subtracting the non-negative parts $E^{2}(\,{\rm or}\thinspace\thinspace F^{2})+G^{2}+H^{2}$
from $A$ (or $C$), the remaining terms represented by $K$ can be
negative or positive, but $|K|$ can not be too large if $G^{2}$
and $H^{2}$ is fixed. The limitation from $G^{2}$ and $H^{2}$
is that $K$ plus $G^{2}$ and $H^{2}$ should be non-negative as
well, since it can be written as a sum of squared forms\footnote{This actually can be understood physically. Cross sections
are generated from squared amplitudes, which makes them always non-negative.
If there are only tensor and scalar interactions, i.e.\ $E=F=0$, the
cross sections should be non-negative as well. Then from the neutrino
cross section (\ref{eq:dm-4}) we have $A\geq0$ by taking the limit
$T\rightarrow E_{\nu}$. A similar argument also applies for $C\geq0$
from the antineutrino cross section. Therefore, without explicit
calculation one can expect that $G^{2}+H^{2}\pm K\geq0$.}
\begin{equation}
G^{2}+H^{2}\pm K=\left(\frac{C_{S}+C_{P}}{4}\right)^{2}+\left(\pm\frac{C_{S}-C_{P}}{4}+C_{T}\right)^{2}+(C\rightarrow D)\,.\label{eq:dm-25}
\end{equation}
This gives the upper bound of $|K|$. On the other hand, there is
no lower bound on $|K|$ because for any given values of $G^{2}$
and $H^{2}$, $K$ always can reach 0 by the cancellation of $C_{S}$
($D_{S}$) with $C_{P}$ ($D_{P}$) in Eq.\ (\ref{eq:dm-24}). Therefore
$|K|$ can be any values from $0$ to $G^{2}+H^{2}$. Based on this
conclusion, we further write
\begin{equation}
K=2GH\cos\gamma\,\label{eq:dm-26}
\end{equation}
which can be considered as the definition of $\gamma$. The advantage
of $\gamma$ is that it can be treated as a free parameter so that
$(A,\thinspace B,\thinspace C)$ can be expressed in terms of  five
free parameters $(E,\thinspace F,\thinspace G,\thinspace H,\thinspace\gamma)$.
As one can check, for any values of $(E,\thinspace F,\thinspace G,\thinspace H,\thinspace\gamma)$,
the corresponding $(C_{a},\thinspace D_{a})$ always exist.

Next we will prove that if
\begin{equation}
L_{\pm}^{2}\equiv G^{2}+H^{2}\pm2GH\cos\gamma\,,\label{eq:dm-27}
\end{equation}
are fixed at any two non-negative values (assuming $L_{\pm}\geq0$
without loss of generality), then
\begin{equation}
-L_{+}L_{-}\leq H^{2}-G^{2}\leq L_{+}L_{-}\,.\label{eq:dm-28}
\end{equation}
To prove Eq.\ (\ref{eq:dm-28}), we start by constructing a parallelogram
with its diagonal lengths equal to $G$ and $H$ and the angle between
them equal to $\gamma$. According to Eq.\ (\ref{eq:dm-27}), the
side lengths of the parallelogram should be $L_{+}$ and $L_{-}$.
If one angle of the parallelogram is defined as $\alpha$, then we
have the following transformation from $(L_{+},\thinspace L_{-},\thinspace\alpha)$
to $(G,\thinspace H,\thinspace\gamma)$:
\begin{eqnarray}
4H^{2} & = & L_{+}^{2}+L_{-}^{2}+2L_{+}L_{-}\cos\alpha\,,\label{eq:dm-29}\\
4G^{2} & = & L_{+}^{2}+L_{-}^{2}-2L_{+}L_{-}\cos\alpha\,,\label{eq:dm-30}\\
\cos\gamma & = & \frac{L_{-}^{2}-L_{+}^{2}}{4HG}\,,\label{eq:dm-31}
\end{eqnarray}
which is essentially a transformation between a parallelogram and
its Varignon parallelogram. From Eqs.\ (\ref{eq:dm-29})
and (\ref{eq:dm-30}) we have
\begin{equation}
H^{2}-G^{2}=L_{+}L_{-}\cos\alpha\,,\label{eq:dm-32}
\end{equation}
which implies that the maximal and minimal values of $H^{2}-G^{2}$
appear when the parallelogram collapses to a line. So $H^{2}-G^{2}$
can be any values from $-L_{+}L_{-}$ to $L_{+}L_{-}$.

From Eqs.\ (\ref{eq:dm-27}), (\ref{eq:dm-28}), (\ref{eq:dm-19})
and (\ref{eq:dm-21}) it is straightforward to derive
\begin{equation}
-\sqrt{A-E^{2}}\sqrt{C-F^{2}}\leq B\leq\sqrt{A-E^{2}}\sqrt{C-F^{2}}\,,\label{eq:dm-33}
\end{equation}
and then
\begin{equation}
B^{2}\leq AC\,.\label{eq:dm-34}
\end{equation}
Converting Eq.\ (\ref{eq:dm-34}) to the corresponding constraint
on the $X-Y$ plane, we get the result in Eq.\ (\ref{eq:dm-17}).

The Majorana bound is much simpler to derive. Since for Majorana neutrinos
$C_{V}=D_{V}=C_{T}=D_{T}=0$, we have
\begin{eqnarray}
A & = & E^{2}+G^{2}\,,\label{eq:dm-19-1}\\
B & = & -G^{2}\,,\label{eq:dm-20-1}\\
C & = & F^{2}+G^{2}\,.\label{eq:dm-21-1}
\end{eqnarray}
Thus, the upper bound of $B$ is 0 while the lower bound depends on $A$
and $C$:
\begin{eqnarray}
A+B & \geq & 0\,,\label{eq:dm-35}\\
C+B & \geq & 0\,.\label{eq:dm-36}
\end{eqnarray}
Then by converting $(A,\thinspace B,\thinspace C)$ to $(X,\thinspace Y)$,
it is straightforward to get the Majorana bound (\ref{eq:dm-18}).

\bibliographystyle{apsrev4-1}
\bibliography{ref}

\end{document}